\documentclass[hidelinks,runningheads]{llncs}
\usepackage{tabularx}
\usepackage{amsmath}
\usepackage{amssymb}
\usepackage{graphicx}
\usepackage{cite}
\usepackage{xcolor}
\usepackage{arydshln}
\usepackage{multirow}
\usepackage{hyperref}
\usepackage{tikz}
\def\checkmark{\tikz\fill[scale=0.4](0,.35) -- (.25,0) -- (1,.7) -- (.25,.15) -- cycle;} 

\usepackage[acronym,shortcuts]{glossaries}
\newacronym{AMD}{AMD}{age-related macular degeneration}
\newacronym{CBAM}{CBAM}{convolutional block attention module}
\newacronym{CNN}{CNN}{convolutional neural network}
\newacronym{DC}{DC}{Dice coefficient}
\newacronym{DR}{DR}{diabetic retinopathy}
\newacronym{ED}{ED}{Euclidean distance}
\newacronym{FCN}{FCN}{fully convolutional network}
\newacronym{HBA}{HBA}{hierarchical bottleneck attention}
\newacronym{mIoU}{mIoU}{mean intersection over union}
\newacronym{MHSA}{MHSA}{multi-head self-attention}
\newacronym{MLP}{MLP}{multi-layer perceptron}
\newacronym{MSE}{MSE}{mean squared error}
\newacronym{NLP}{NLP}{natural language processing}
\newacronym{OD}{OD}{optic disc}
\newacronym{R-CNN}{R-CNN}{region-based convolutional neural network}
\newacronym{ROI}{ROI}{region of interest}
\newacronym{scSE}{scSE}{concurrent spatial and channel squeeze \& excitation}
\newacronym{SOTA}{SOTA}{state-of-the-art}

\begin{document}
%
\title{
U-Net with Hierarchical Bottleneck Attention for Landmark Detection in Fundus Images \\ of the Degenerated Retina}

\titlerunning{Joint Fovea and OD Segmentation for Retinal Degeneration}
%
\author{Shuyun Tang\inst{1} \and
Ziming Qi\inst{1} \and
Jacob Granley\inst{1} \and
Michael Beyeler\inst{1}}
\authorrunning{Tang et al.}
%
\institute{University of California, Santa Barbara, CA 93106
\email{\{shuyun,zimingqi,jgranley,mbeyeler\}@ucsb.edu}}
\maketitle  
\begin{abstract}
Fundus photography has routinely been used to document the presence and severity of retinal degenerative diseases such as \acf{AMD}, glaucoma, and \acf{DR} in clinical practice, for which the fovea and \acf{OD} are important retinal landmarks.
However, the occurrence of lesions, drusen, and other retinal abnormalities during retinal degeneration severely complicates automatic landmark detection and segmentation.
Here we propose HBA-U-Net: a U-Net backbone enriched with hierarchical bottleneck attention. The network consists of a novel bottleneck attention block that combines and refines self-attention, channel attention, and relative-position attention to highlight retinal abnormalities that may be important for fovea and OD segmentation in the degenerated retina.
HBA-U-Net achieved state-of-the-art results on fovea detection across datasets and eye conditions (ADAM: \acf{ED} of 25.4 pixels, REFUGE: 32.5 pixels, IDRiD: 32.1 pixels), on OD segmentation for AMD (ADAM: \acf{DC} of 0.947), and on OD detection for DR (IDRiD: ED of 20.5 pixels).
We further validated the design of our network with an ablation study.
Our results suggest that HBA-U-Net may be well suited for landmark detection in the presence of a variety of retinal degenerative diseases.

\end{abstract}
\section{Introduction}

\Ac{AMD}, glaucoma, and \ac{DR} are three of the most common causes of blindness in the world \cite{steinmetz_causes_2021}.
Fundus photography has routinely been used to document the presence and severity of these retinal degenerative diseases in clinical practice.
Among the landmarks of interest are the fovea, which is a small depression in the macula, and the \ac{OD}, which is where the optic nerve and blood vessels leave the retina.
However, detecting retinal abnormalities associated with these diseases (e.g.,  drusen in \ac{AMD}, hemorrhage in \ac{DR}) is a labor-intensive and time-consuming process, thus necessitating the need for automated fundus image analysis.

In recent years, numerous methods have been proposed for retinal structure detection.
Jiang \emph{et al.} \cite{jiang_robust_2019} proposed an encoder-decoder network with deep residual structure and recursive learning mechanism for robust \ac{OD} localization, followed by an end-to-end \ac{R-CNN} for joint optic disc and cup segmentation \cite{jiang_jointrcnn_2020}.
Similarly, numerous studies have employed various \ac{CNN} models for fovea localization (e.g., \cite{sedai_multi-stage_2017,alais_fast_2020}).
Although fovea and \ac{OD} are spatially correlated with each other, only a few studies (e.g., \cite{yu_fast_2011,meyer_pixel-wise_2018}) have focused on joint fovea and \ac{OD} segmentation.
Furthermore, models trained on healthy eyes tend not to generalize well to diseased eyes with retinal abnormalities. A notable exception is Kamble \emph{et al.} \cite{kamble_optic_2020} who achieved \ac{SOTA} performance on landmark detection for \ac{AMD} and glaucoma using a modified U-Net\verb!++! with an EfficientNet encoder.
However, there is potential merit in combining convolutional backbone networks with attentional mechanisms \cite{vaswani_attention_2017} to highlight retinal abnormalities that may be important for landmark detection in the degenerated retina.

To develop a segmentation model that is well suited for retinal degeneration, 
we propose HBA-U-Net: a U-Net backbone enriched with hierarchical bottleneck attention. 
The main contributions of this work are:
\begin{enumerate}
    \item We propose a \ac{HBA} block: a novel attention mechanism that combines and refines self-attention \cite{vaswani_attention_2017}, channel attention \cite{DBLP:journals/corr/abs-1807-06521}, and relative-position attention \cite{ramachandran_stand-alone_2019} to highlight retinal abnormalities important for landmark detection in the degenerated retina.
    \item We integrate the \ac{HBA} block into bottleneck skip connections across all layers of a U-Net backbone network to form HBA-U-Net, and test the network's performance on three benchmark datasets for retinal degeneration: ADAM \cite{fu_adam_2020} for \ac{AMD}, REFUGE \cite{ORLANDO2020101570} for glaucoma, and IDRiD \cite{PORWAL2020101561} for \ac{DR}.
    \item We validate the design of HBA-U-Net with an ablation study.
    \item We demonstrate \ac{SOTA} performance on fovea detection across datasets and eye conditions, on \ac{OD} segmentation for \ac{AMD}, and on \ac{OD} detection for \ac{DR}.
\end{enumerate}

\section{Methods}

\subsection{Model Architecture}

\subsubsection{HBA-U-Net}
\begin{figure}[!t]
    \centering
    \includegraphics[width=\textwidth]{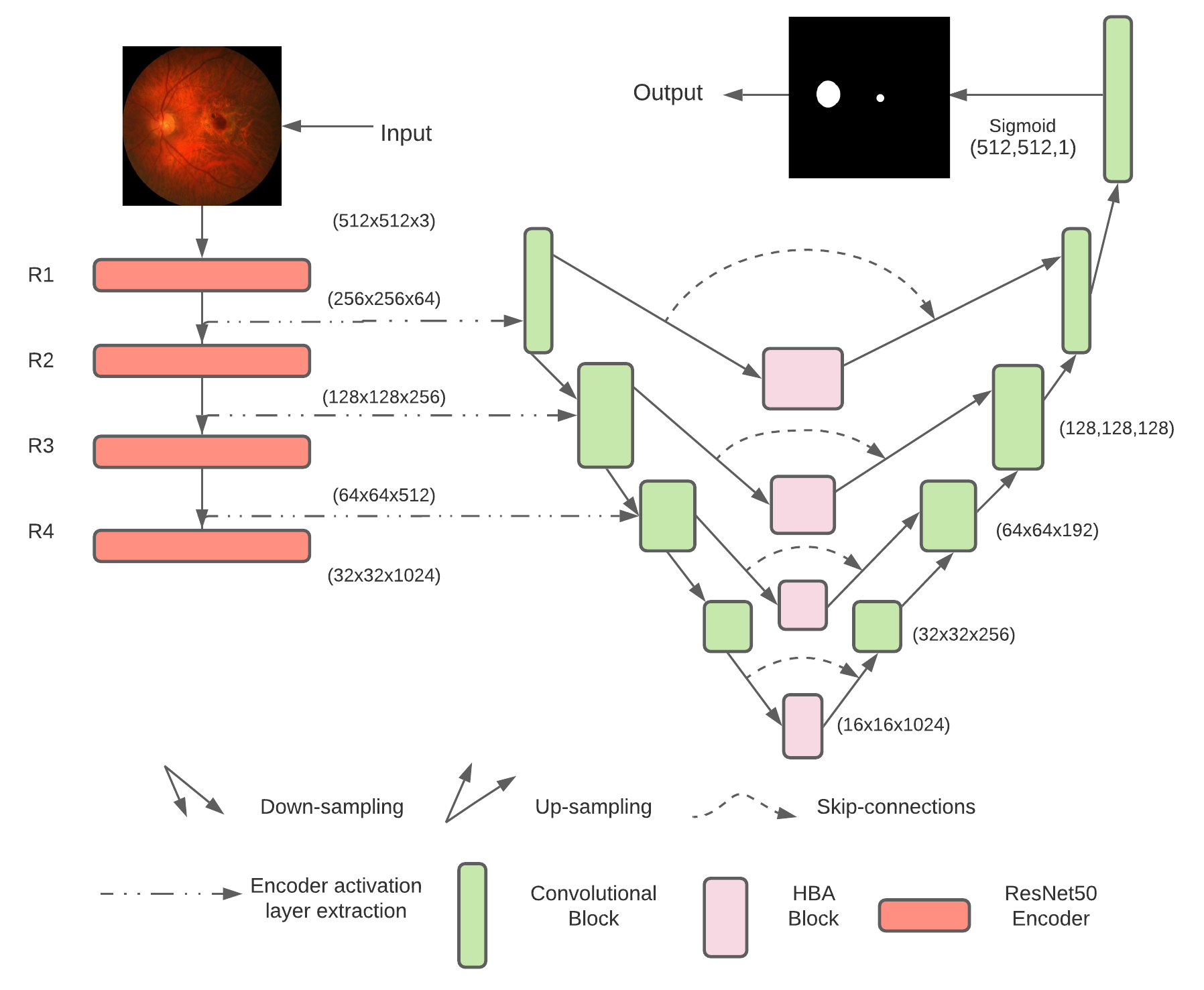}
    \caption{HBA-U-Net architecture. A U-Net enriched with a novel attention block and re-designed skip-connection paths jointly locates the fovea and segments the optic disc. ResNet-50 was used as encoder. Note that the number of local bottlenecks and the down/up-sampling projection rate depends on the image dimensions. }
    \label{fig:architecture}
\end{figure}

The proposed network architecture is illustrated in Fig.~\ref{fig:architecture}. 
First, ImageNet pretrained ResNet-50 blocks were used as encoders to obtain feature maps at different spatial resolutions. These feature maps, along with the original image, were then fed into a modified U-Net structure \cite{ronneberger_u-net_2015,meyer_pixel-wise_2018} with \ac{HBA} blocks added to skip connections.
The outputs of the \ac{HBA} blocks were up-sampled and aggregated to produce the final fovea and \ac{OD} segmentation mask. 

Our goal was to incorporate \ac{HBA} blocks into the U-Net without drastically increasing the computational complexity.
Consistent with \cite{srinivas_bottleneck_2021}, we noticed that adding a self-attention mechanism to the bottleneck layers (a shrinking path, the attention module, and an expanding path) significantly boosted the network's performance.
However, the original U-Net contains only a single bottleneck layer (between the last down-sampling block and the first up-sampling block).
To incorporate multiple HBA blocks into the network, we therefore re-designed the U-Net by creating local bottleneck structures in each skip-connection pair (see Fig.~\ref{fig:architecture}).
After each down-convolution block, the features were down-sampled by pooling and passed to the HBA block, followed by up-sampling to the original size. In this way, the pairs of down/up-sampling convolution blocks could be treated as local bottleneck structures operating at different spatial resolutions. 

\subsubsection{HBA block}

\begin{figure}[!t]
    \centering
    \includegraphics[width=\textwidth, scale=0.4]{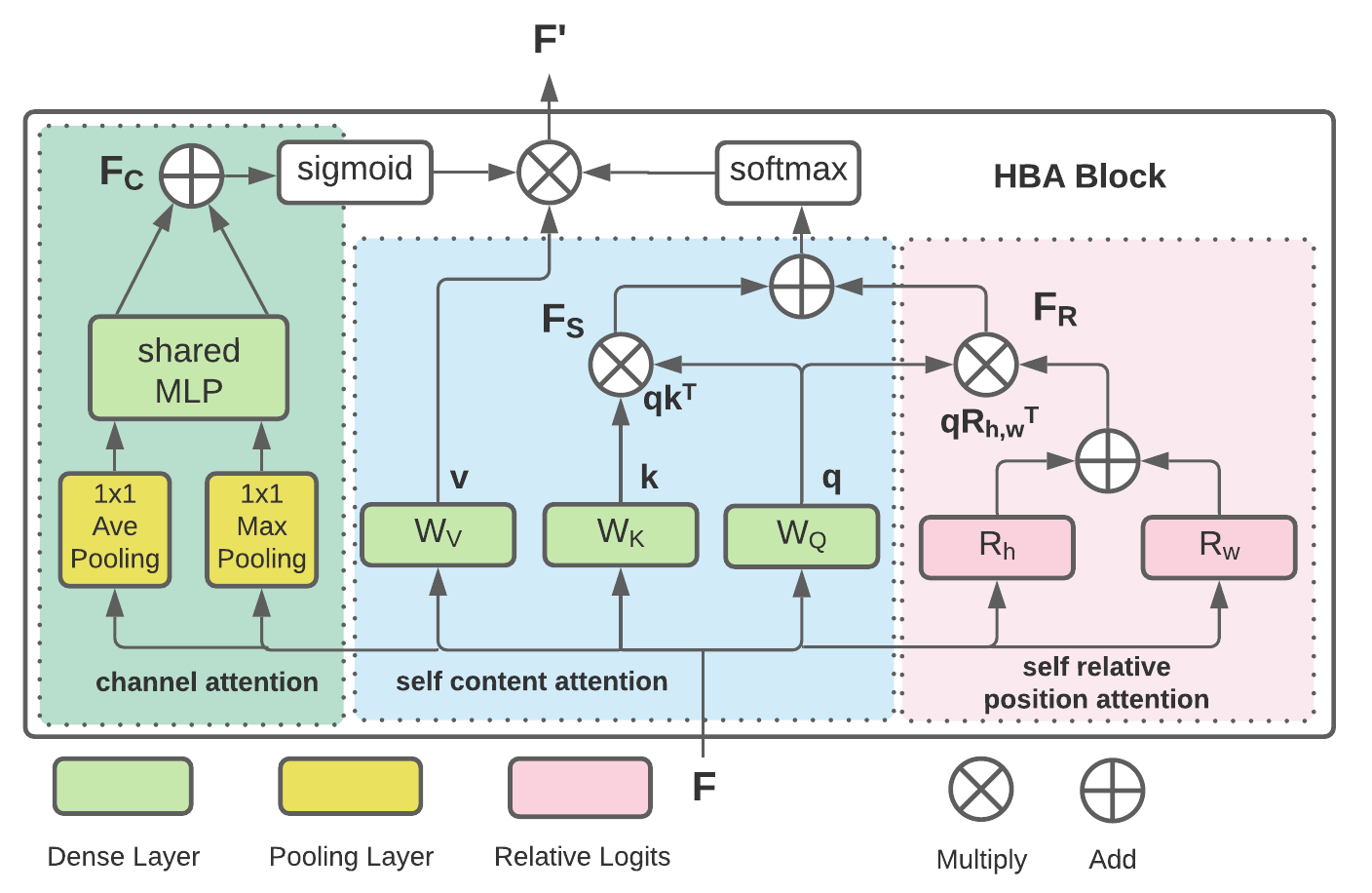}
    \caption{HBA block architecture, consisting of channel attention (green box outputs $F_C$), content attention using multi-head self-attention (blue box outputs $F_S$), and relative position attention (pink box outputs $F_R$).}
    \label{fig:MHSA-block}
\end{figure}

Recently, attention mechanisms have seen widespread adoption in various tasks \cite{vaswani_attention_2017}. 
Inspired by \cite{srinivas_bottleneck_2021,DBLP:journals/corr/abs-1807-06521}, our \ac{HBA} block (Fig.~\ref{fig:MHSA-block}) consisted of channel, content, and relative-position attention modules, each described in detail below. We denote the query, key, value, input feature map, relative height logit, and relative weight logit as $q, k, v, F, R_h, R_w$, respectively. 

In the proposed \ac{HBA} block, content attention (blue box in Fig.~\ref{fig:MHSA-block}) attended to individual pixels in each spatial feature map. For each attention head, dense layers ($W_Q, W_K, W_V$) were used to calculate the query ($q = W_Q(F)$), key ($k = W_K(F)$), and value ($v = W_V(F)$) for each pixel. The output of the content attention was an attention score ($F_S$) between key ($k$) and query vectors ($q$):
\begin{equation} \label{eq:content}
    \mathrm{F_S} = qk^T.
\end{equation}

Inspired by \cite{ramachandran_stand-alone_2019,srinivas_bottleneck_2021}, we included relative-position attention (pink box in Fig.~\ref{fig:MHSA-block}) to encode the relative position of different retinal landmarks (e.g., to relate the fovea to the \ac{OD} location).
Relative logits were used to store the $x$ and $y$ offsets ($R_h$ and $R_w$) between each key and query. These were added and the relative positional attention score $\mathrm{F_R}$ was computed using the dot product:
\begin{equation} \label{eq:position}
    \mathrm{F_R} = q(R_h+R_w)^T.
\end{equation}
  


In a U-Net, spatial information is encoded to different channels through down/up-sampling. We believe channel-wise attention is well suited to utilize this information in the bottleneck layers, which usually have many channels. 
We therefore used the channel attention module proposed in \cite{DBLP:journals/corr/abs-1807-06521} (green box in Fig. \ref{fig:MHSA-block}).
The input feature map $\mathrm{F}$
was passed in parallel to average pooling and max pooling layers, compressing each channel to one value.
These two feature maps were forwarded through a single, shared \ac{MLP} with one hidden layer and added to compute the final channel attention score ($F_C$):
\begin{equation} \label{eq:channel}
    \mathrm{F_C} = MLP(AvgPool(F)) + MLP(MaxPool(F)).
\end{equation}

In contrast to a conventional transformer, the value vector was scaled not only by the content attention score ($F_S$), but also according to the relative-position attention score ($F_R$) and the channel attention score ($F_C$). 
The output of the \Ac{HBA} block ($\mathrm{F'}$) is given as follows:
\begin{equation} \label{eq:hba}
    \mathrm{F'} = softmax(F_S+F_R)\sigma(F_C)v,
\end{equation}
where the softmax was applied across attention heads and $\sigma$ denotes the sigmoid function.

\subsection{Datasets}

We evaluated our model on three prominent datasets for retinal degeneration: ADAM \cite{fu_adam_2020} for \ac{AMD}, REFUGE \cite{ORLANDO2020101570} for glaucoma, and IDRiD \cite{PORWAL2020101561} for \ac{DR}.

ADAM was released as part of a Grand Challenge at a satellite event of the ISBI 2020 conference.
The dataset contains 400 fundus images at either $2124 \times 2056$ or $1444 \times 1444$ resolution, 87 of which depict eyes at various stages of \ac{AMD} progression (typical signs include the presence of drusen, exudation, and hemorrhage), and the rest are from healthy controls.
ADAM includes ground-truth \ac{OD} segmentation masks and fovea image coordinates.

REFUGE was released as part of a Grand Challenge of the OMIA5 workshop at MICCAI 2018. The dataset contains 1200 fundus images at either $2124 \times 2056$ or $1634 \times 1634$  resolution, 120 of which depict eyes with glaucoma, and the rest are from healthy controls.
REFUGE includes ground-truth \ac{OD} segmentation masks and fovea image coordinates.

IDRiD was released as part of a Grand Challenge at ISBI 2018.
The dataset contains 516 images at $4288 \times 2848$ resolution divided into 413 train images and 103 test images, all of which contain pathological conditions such as \ac{DR} and diabetic macular edema.
IDRiD includes ground-truth image coordinates for the fovea and \ac{OD} center, but not segmentation masks.

\subsection{Implementation Details}

\subsubsection{Data Preprocessing and Augmentation}
First, we resized every image in the dataset to $512 \times 512$ pixels.
Second, we followed \cite{kamble_optic_2020} to generate circular segmentation masks from the ground-truth fovea coordinates and combined them with the ground-truth \ac{OD} segmentation masks.
Third, we applied random image rotations (uniformly sampled from $[-0.2, 0.2]$ rad), and horizontal/vertical flips to augment the original dataset on-the-fly. Fourth, we split the data 85-15 into train and test sets and held out 20\% of the training images for validation.

\subsubsection{Training Procedure}
The model was trained using the adam optimizer, the Dice loss \cite{sudre_generalised_2017}, and early stopping, with a custom learning rate scheduler (start rate 0.0025, decay rate 0.985 after 150 epochs), and batch size 8 for 500 epochs.
Initial weights were pre-trained on ImageNet.
The model was implemented using Keras 2.4.3 (Python 3.7) and run on an NVIDIA Tesla K80 (12GB of RAM) provided by Google Colab Pro.
The code is available at \url{github.com/anonymous}.

\subsubsection{Evaluation Metrics}

We evaluated the performance of the model using \ac{ED} \cite{meyer_pixel-wise_2018}, where only image coordinates were given, and \ac{DC}, where segmentation masks were given.
Since none of the three datasets came with fovea segmentation masks, we followed \cite{meyer_pixel-wise_2018} to create a circular disc centered over the ground-truth fovea coordinates, which was then used to train our network.
After training, we recovered predicted coordinates by extracting the centroid of the predicted segmentation mask using scikit-image.




\section{Experiments and Results}

\begin{table}[!b]
    \centering
    \setlength\tabcolsep{4pt}
    \def\arraystretch{1.2}%
    \caption{Landmark detection on ADAM, REFUGE, and IDRiD. Note that Challenge test data is not publicly available. ED: Euclidean Distance. DC: Dice Coefficient. }
    \begin{tabularx}{\textwidth}{l|X|r|rr}
        \multicolumn{2}{l|}{} & Fovea & \multicolumn{2}{c}{Optic Disc} \\
        \hline
        & Model  & ED & ED & DC \\
        \hline
        \multirow{3}{*}[1.2ex]{\rotatebox{90}{ {\scriptsize ADAM} }} &
        Aira matrix \cite{kamble_optic_2020} {\scriptsize (ISBI 2020 Challenge Winner)} & 26.2 & - & {\bf 0.947}  \\
        & HBA-U-Net (this paper) & {\bf 25.4} & - & {\bf 0.947}  \\
        \hline
        \multirow{5}{*}{\rotatebox{90}{ {\scriptsize REFUGE} }} & Fu \emph{et al.} \cite{fu_disc-aware_2018} & - & - & 0.936 \\
        & Zhang \emph{et al.} \cite{zhang_et-net_2019} & - & - & 0.953  \\
        & Kamble \emph{et al.} \cite{kamble_optic_2020} &  35.2 & - & 0.957  \\
        & Wang \emph{et al.} \cite{wang_patch-based_2019} & - & - & {\bf 0.960}  \\
        & HBA-U-Net (this paper) & {\bf 32.5} & - & 0.947   \\
        \hline
        \multirow{3}{*}{\rotatebox{90}{ {\scriptsize IDRiD} }}
        & DeepDR {\scriptsize (IDRiD Subchallenge-3 Winner, on-site)} & 64.5 & 21.1 & -  \\
        & ZJU-BII-SGEX {\scriptsize (IDRiD Subchallenge-3 Winner, online)} & 45.9 & 25.6 & -  \\
        & HBA-U-Net (this paper) &  {\bf 32.1} & {\bf 20.5} & -  \\
    \end{tabularx}
    \label{tab:sota}
\end{table}

\subsection{Joint Fovea and OD Detection in the Degenerated Retina}
Table~\ref{tab:sota} summarizes our results on three prominent datasets for retinal degeneration: ADAM for \ac{AMD}, REFUGE for glaucoma, and IDRiD for \ac{DR}.

HBA-U-Net achieved \ac{SOTA} performance on fovea detection across all datasets  (ADAM: \ac{ED} 25.4 px; REFUGE: \ac{ED} 32.5 px; IDRiD: 32.1 px) and thus across eye conditions, despite the fact that these datasets were previously used in Grand Challenges that featured convolutional \cite{kamble_optic_2020}, attentional \cite{zhang_et-net_2019}, and adversarial \cite{wang_patch-based_2019} approaches, some of which had a considerably larger number of trainable parameters.
Because all three datasets are relatively new, the number of published results is still relatively small.

HBA-U-Net also achieved \ac{SOTA} performance on \ac{OD} segmentation for \ac{AMD} (\ac{DC} of 0.947, on par with \cite{kamble_optic_2020}) and on \ac{OD} detection for \ac{DR} (\ac{ED} of 20.5).
Our \ac{OD} segmentation was slightly worse than competing models, with the \ac{SOTA} belonging to \cite{wang_patch-based_2019}, a patch-based morphology-aware segmentation network.

\begin{figure}[!t]
    \centering
    \includegraphics[width=\textwidth]{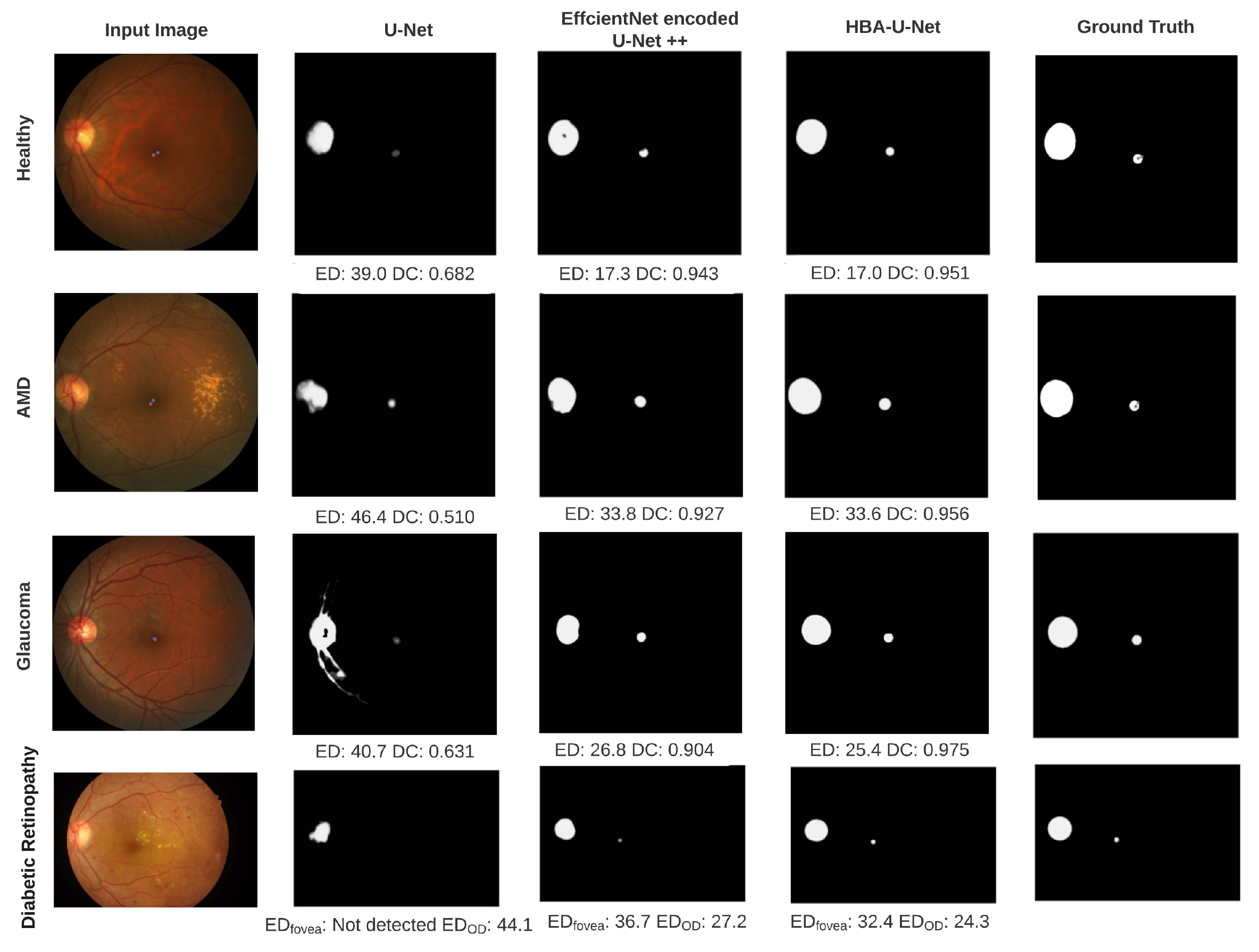}
    \caption{Representative example predictions for a healthy eye (\emph{top row}), \ac{AMD} (\emph{second row}), glaucoma (\emph{third row}), and \ac{DR} (\emph{bottom row}).
    Predictions are shown for a re-implemented U-Net (\emph{second column}), EfficientNet encoded U-Net\texttt{++} with scSE blocks (\emph{third column}), and HBA-U-Net (\emph{fourth column}), and compared against ground truth (\emph{rightmost column}). Error rates are given below each prediction panel.
    }
    \label{fig:example-predictions}
\end{figure}

However, please note that the test data of these challenges is not made available to the public.
To offer a fair comparison across models, we therefore re-implemented a number of commonly used alternative network architectures and compared their performance using our own train/test split. These alternative networks included the classical U-Net \cite{ronneberger_u-net_2015} and an EfficientNet \cite{tan_efficientnet_2020} encoded U-Net\verb!++! with scSE blocks 
(similar to \cite{kamble_optic_2020}). Results are given in Table \ref{tab:reimplemented_models} and example predictions are shown in Fig.~\ref{fig:example-predictions}. HBA-U-Net outperformed the baseline models on all three datasets.


\begin{table}[t!]
    \caption{Landmark detection for different reimplemented models tested on ADAM, REFUGE, and IDRiD.
    ED: Euclidean Distance, DC: Dice Coefficient, F: Fovea, OD: Optic Disc.}
    \centering
    \setlength\tabcolsep{4pt}
    \def\arraystretch{1.2}%
    \begin{tabularx}{\textwidth}{X|cc|cc|cc}
    &  \multicolumn{2}{c|}{ADAM} & \multicolumn{2}{c|}{REFUGE} & \multicolumn{2}{c}{IDRiD} \\
    Model &  $\mathrm{ED}_\mathrm{F}$ & $\mathrm{DC}_\mathrm{OD}$ & $\mathrm{ED}_\mathrm{F}$ & $\mathrm{DC}_\mathrm{OD}$ & $\mathrm{ED}_\mathrm{F}$ & $\mathrm{ED}_\mathrm{OD}$ \\ \hline
    U-Net \cite{ronneberger_u-net_2015}  & 70.7 & 0.741 &  65.2 &  0.806&  87.1&   53.7\\
    EfficientNet encoded U-Net\texttt{++} \cite{kamble_optic_2020}
    & 26.9 & 0.867 &  37.6&  0.935&  50.4&  28.1\\
    HBA-U-Net (this paper) & {\bf 25.4} & {\bf 0.947}  &  {\bf 32.5}& {\bf 0.947}& {\bf 32.1}& {\bf 20.5} 
    \end{tabularx}
    \label{tab:reimplemented_models}
\end{table}


\subsection{Ablation Study}

To measure the impact of the HBA block on different versions of our proposed model architecture, we performed an ablation study on ADAM (see Table~\ref{tab:adam-ablation}).

Starting with the original U-Net \cite{ronneberger_u-net_2015,meyer_pixel-wise_2018} as a baseline, we were able to reduce fovea \ac{ED} by a factor of two by adding a ResNet-50 encoder \cite{he_deep_2016}.
Adding the original self-attention block \cite{ramachandran_stand-alone_2019} (without relative position and channel-wise attention; labeled `Self-Att' in Table~\ref{tab:adam-ablation}) at the bottleneck part of the U-Net improved fovea \ac{ED} by $\sim 5\%$, but led to a $\sim 2\%$ decrease in \ac{DC} for \ac{OD} segmentation. 
Upgrading the self-attention block to our proposed HBA block at the bottleneck part of the U-Net (labeled `HBA-1') resulted in both the \ac{ED} and \ac{DC} improving by $\sim 4\%$. Finally, creating local bottlenecks with HBA blocks at each skip connection in the hierarchy (labeled `HBA-all') led to \ac{SOTA} performance.


\begin{table}[!t]
    \centering
    \caption{Ablation studies on each network component. Starting from a U-Net backbone~\cite{ronneberger_u-net_2015,meyer_pixel-wise_2018}, we gradually added a ResNet-50 encoder~\cite{he_deep_2016}, a standard self-attention block~\cite{ramachandran_stand-alone_2019} (`Self-Att'), a single HBA block in the bottleneck (`HBA-1'), and HBA blocks across all levels of the hierarchy (`HBA-all').}
    \setlength\tabcolsep{4pt}
    \def\arraystretch{1.2}%
    \begin{tabularx}{\textwidth}{cccccXr|r|r}
        \hline
        U-Net & ResNet & Self-Att & HBA-1 & HBA-all & & Params &  Fovea ED & OD DC \\
        \hline
        \checkmark & & & & & & 8.7M &  70.7 & 0.741 \\
        \checkmark & \checkmark & & & & & 20.8M &  34.8 & 0.902  \\
        \checkmark & \checkmark & \checkmark & & & & 21.1M &  29.8 & 0.925 \\
        \checkmark & \checkmark & \checkmark & \checkmark & & & 21.3M &  25.8 & 0.920 \\
        \checkmark & \checkmark & \checkmark & \checkmark & \checkmark & & 22.2M & {\bf 25.4} & {\bf 0.947}  \\
        \hline
    \end{tabularx}
    \label{tab:adam-ablation}
\end{table}

\section{Conclusions}

We have proposed a re-designed U-Net architecture with hierarchical bottleneck attention and demonstrated its utility for fundus analysis.
The proposed network achieved \ac{SOTA} performance on fovea detection across datasets and eye conditions, on \ac{OD} segmentation for \ac{AMD}, and on \ac{OD} detection for \ac{DR}.

Although self-attention, channel attention, and relative-position have been deployed separately in other computer vision tasks, here we refined, simplified, and combined their potential in segmenting retinal abnormalities. 
Furthermore, our ablation study demonstrates the benefit of the local bottleneck structures and \ac{HBA} blocks
for retinal landmark segmentation.
Compared to content self-attention alone, \ac{HBA} does not add much overhead: relative position attention does not have any learnable parameters and channel attention consists of a shared \ac{MLP} with one hidden layer.
Compared to other pure attention networks such as ViT \cite{dosovitskiy2021image}, \ac{HBA} blocks are more resourceful and better suited to work in combination with convolutional modules commonly used in segmentation tasks.

Overall our results suggest that HBA-U-Net may be well suited for landmark detection in the presence of a variety of retinal degenerative diseases.

\newpage

%
%
\bibliographystyle{splncs04}
\bibliography{references}

\begin{thebibliography}{10}
\providecommand{\url}[1]{\texttt{#1}}
\providecommand{\urlprefix}{URL }
\providecommand{\doi}[1]{https://doi.org/#1}

\bibitem{alais_fast_2020}
Alais, R., Dokládal, P., Erginay, A., Figliuzzi, B., Decencière, E.: Fast
  macula detection and application to retinal image quality assessment.
  Biomedical Signal Processing and Control  \textbf{55},  101567 (Jan 2020)

\bibitem{steinmetz_causes_2021}
Blindness, G.., Collaborators, V.I.: Causes of blindness and vision impairment
  in 2020 and trends over 30 years, and prevalence of avoidable blindness in
  relation to {VISION} 2020: the {Right} to {Sight}: an analysis for the
  {Global} {Burden} of {Disease} {Study}. The Lancet Global Health
  \textbf{9}(2),  e144--e160 (Feb 2021), publisher: Elsevier

\bibitem{dosovitskiy2021image}
Dosovitskiy, A., Beyer, L., Kolesnikov, A., Weissenborn, D., Zhai, X.,
  Unterthiner, T., Dehghani, M., Minderer, M., Heigold, G., Gelly, S.,
  Uszkoreit, J., Houlsby, N.: An image is worth 16x16 words: Transformers for
  image recognition at scale (2021)

\bibitem{fu_adam_2020}
Fu, H.: {ADAM}: {Automatic} {Detection} challenge on {Age}-related {Macular}
  degeneration (Jan 2020), publisher: IEEE type: dataset

\bibitem{fu_disc-aware_2018}
Fu, H., Cheng, J., Xu, Y., Zhang, C., Wong, D.W.K., Liu, J., Cao, X.:
  Disc-{Aware} {Ensemble} {Network} for {Glaucoma} {Screening} {From} {Fundus}
  {Image}. IEEE Transactions on Medical Imaging  \textbf{37}(11),  2493--2501
  (Nov 2018), conference Name: IEEE Transactions on Medical Imaging

\bibitem{he_deep_2016}
He, K., Zhang, X., Ren, S., Sun, J.: Deep {Residual} {Learning} for {Image}
  {Recognition}. In: 2016 {IEEE} {Conference} on {Computer} {Vision} and
  {Pattern} {Recognition} ({CVPR}). pp. 770--778 (Jun 2016), iSSN: 1063-6919

\bibitem{jiang_robust_2019}
Jiang, S., Chen, Z., Li, A., Wang, Y.: Robust {Optic} {Disc} {Localization} by
  {Large} {Scale} {Learning}. In: Fu, H., Garvin, M.K., MacGillivray, T., Xu,
  Y., Zheng, Y. (eds.) Ophthalmic {Medical} {Image} {Analysis}. pp. 95--103.
  Lecture {Notes} in {Computer} {Science}, Springer International Publishing,
  Cham (2019)

\bibitem{jiang_jointrcnn_2020}
Jiang, Y., Duan, L., Cheng, J., Gu, Z., Xia, H., Fu, H., Li, C., Liu, J.:
  {JointRCNN}: {A} {Region}-{Based} {Convolutional} {Neural} {Network} for
  {Optic} {Disc} and {Cup} {Segmentation}. IEEE Transactions on Biomedical
  Engineering  \textbf{67}(2),  335--343 (Feb 2020), conference Name: IEEE
  Transactions on Biomedical Engineering

\bibitem{kamble_optic_2020}
Kamble, R., Samanta, P., Singhal, N.: Optic {Disc}, {Cup} and {Fovea}
  {Detection} from {Retinal} {Images} {Using} {U}-{Net}++ with {EfficientNet}
  {Encoder}. In: Fu, H., Garvin, M.K., MacGillivray, T., Xu, Y., Zheng, Y.
  (eds.) Ophthalmic {Medical} {Image} {Analysis}. pp. 93--103. Lecture {Notes}
  in {Computer} {Science}, Springer International Publishing, Cham (2020)

\bibitem{meyer_pixel-wise_2018}
Meyer, M.I., Galdran, A., Mendonça, A.M., Campilho, A.: A {Pixel}-{Wise}
  {Distance} {Regression} {Approach} for {Joint} {Retinal} {Optical} {Disc} and
  {Fovea} {Detection}. In: Frangi, A.F., Schnabel, J.A., Davatzikos, C.,
  Alberola-López, C., Fichtinger, G. (eds.) Medical {Image} {Computing} and
  {Computer} {Assisted} {Intervention} – {MICCAI} 2018. pp. 39--47. Lecture
  {Notes} in {Computer} {Science}, Springer International Publishing, Cham
  (2018)

\bibitem{ORLANDO2020101570}
Orlando, J.I., Fu, H., {Barbosa Breda}, J., {van Keer}, K., Bathula, D.R.,
  Diaz-Pinto, A., Fang, R., Heng, P.A., Kim, J., Lee, J., Lee, J., Li, X., Liu,
  P., Lu, S., Murugesan, B., Naranjo, V., Phaye, S.S.R., Shankaranarayana,
  S.M., Sikka, A., Son, J., {van den Hengel}, A., Wang, S., Wu, J., Wu, Z., Xu,
  G., Xu, Y., Yin, P., Li, F., Zhang, X., Xu, Y., Bogunović, H.: Refuge
  challenge: A unified framework for evaluating automated methods for glaucoma
  assessment from fundus photographs. Medical Image Analysis  \textbf{59},
  101570 (2020)

\bibitem{PORWAL2020101561}
Porwal, P., Pachade, S., Kokare, M., Deshmukh, G., Son, J., Bae, W., Liu, L.,
  Wang, J., Liu, X., Gao, L., Wu, T., Xiao, J., Wang, F., Yin, B., Wang, Y.,
  Danala, G., He, L., Choi, Y.H., Lee, Y.C., Jung, S.H., Li, Z., Sui, X., Wu,
  J., Li, X., Zhou, T., Toth, J., Baran, A., Kori, A., Chennamsetty, S.S.,
  Safwan, M., Alex, V., Lyu, X., Cheng, L., Chu, Q., Li, P., Ji, X., Zhang, S.,
  Shen, Y., Dai, L., Saha, O., Sathish, R., Melo, T., Araújo, T., Harangi, B.,
  Sheng, B., Fang, R., Sheet, D., Hajdu, A., Zheng, Y., Mendonça, A.M., Zhang,
  S., Campilho, A., Zheng, B., Shen, D., Giancardo, L., Quellec, G.,
  Mériaudeau, F.: Idrid: Diabetic retinopathy – segmentation and grading
  challenge. Medical Image Analysis  \textbf{59},  101561 (2020)

\bibitem{ramachandran_stand-alone_2019}
Ramachandran, P., Parmar, N., Vaswani, A., Bello, I., Levskaya, A., Shlens, J.:
  Stand-{Alone} {Self}-{Attention} in {Vision} {Models}. arXiv:1906.05909 [cs]
  (Jun 2019), arXiv: 1906.05909

\bibitem{ronneberger_u-net_2015}
Ronneberger, O., Fischer, P., Brox, T.: U-{Net}: {Convolutional} {Networks} for
  {Biomedical} {Image} {Segmentation}. In: Navab, N., Hornegger, J., Wells,
  W.M., Frangi, A.F. (eds.) Medical {Image} {Computing} and
  {Computer}-{Assisted} {Intervention} – {MICCAI} 2015. pp. 234--241. Lecture
  {Notes} in {Computer} {Science}, Springer International Publishing, Cham
  (2015)

\bibitem{sedai_multi-stage_2017}
Sedai, S., Tennakoon, R., Roy, P., Cao, K., Garnavi, R.: Multi-stage
  segmentation of the fovea in retinal fundus images using fully
  {Convolutional} {Neural} {Networks}. In: 2017 {IEEE} 14th {International}
  {Symposium} on {Biomedical} {Imaging} ({ISBI} 2017). pp. 1083--1086 (Apr
  2017), iSSN: 1945-8452

\bibitem{srinivas_bottleneck_2021}
Srinivas, A., Lin, T.Y., Parmar, N., Shlens, J., Abbeel, P., Vaswani, A.:
  Bottleneck {Transformers} for {Visual} {Recognition}. arXiv:2101.11605 [cs]
  (Jan 2021), arXiv: 2101.11605

\bibitem{sudre_generalised_2017}
Sudre, C.H., Li, W., Vercauteren, T., Ourselin, S., Jorge~Cardoso, M.:
  Generalised {Dice} {Overlap} as a {Deep} {Learning} {Loss} {Function} for
  {Highly} {Unbalanced} {Segmentations}. In: Cardoso, M.J., Arbel, T.,
  Carneiro, G., Syeda-Mahmood, T., Tavares, J.M.R., Moradi, M., Bradley, A.,
  Greenspan, H., Papa, J.P., Madabhushi, A., Nascimento, J.C., Cardoso, J.S.,
  Belagiannis, V., Lu, Z. (eds.) Deep {Learning} in {Medical} {Image}
  {Analysis} and {Multimodal} {Learning} for {Clinical} {Decision} {Support}.
  pp. 240--248. Lecture {Notes} in {Computer} {Science}, Springer International
  Publishing, Cham (2017)

\bibitem{tan_efficientnet_2020}
Tan, M., Le, Q.V.: {EfficientNet}: {Rethinking} {Model} {Scaling} for
  {Convolutional} {Neural} {Networks}. arXiv:1905.11946 [cs, stat]  (Sep 2020),
  arXiv: 1905.11946

\bibitem{vaswani_attention_2017}
Vaswani, A., Shazeer, N., Parmar, N., Uszkoreit, J., Jones, L., Gomez, A.N.,
  Kaiser, L., Polosukhin, I.: Attention is {All} you {Need}. Advances in Neural
  Information Processing Systems  \textbf{30} (2017)

\bibitem{wang_patch-based_2019}
Wang, S., Yu, L., Yang, X., Fu, C.W., Heng, P.A.: Patch-{Based} {Output}
  {Space} {Adversarial} {Learning} for {Joint} {Optic} {Disc} and {Cup}
  {Segmentation}. IEEE Transactions on Medical Imaging  \textbf{38}(11),
  2485--2495 (Nov 2019), conference Name: IEEE Transactions on Medical Imaging

\bibitem{DBLP:journals/corr/abs-1807-06521}
Woo, S., Park, J., Lee, J., Kweon, I.S.: {CBAM:} convolutional block attention
  module. CoRR  (2018)

\bibitem{yu_fast_2011}
Yu, H., Barriga, S., Agurto, C., Echegaray, S., Pattichis, M., Zamora, G.,
  Bauman, W., Soliz, P.: Fast localization of optic disc and fovea in retinal
  images for eye disease screening. In: Medical {Image} {Computing} and
  {Computer} {Assisted} {Intervention} – {MICCAI} 2011. vol.~7963, p. 796317
  (Mar 2011)

\bibitem{zhang_et-net_2019}
Zhang, Z., Fu, H., Dai, H., Shen, J., Pang, Y., Shao, L.: {ET}-{Net}: {A}
  {Generic} {Edge}-{aTtention} {Guidance} {Network} for {Medical} {Image}
  {Segmentation}. In: Shen, D., Liu, T., Peters, T.M., Staib, L.H., Essert, C.,
  Zhou, S., Yap, P.T., Khan, A. (eds.) Medical {Image} {Computing} and
  {Computer} {Assisted} {Intervention} – {MICCAI} 2019. pp. 442--450. Lecture
  {Notes} in {Computer} {Science}, Springer International Publishing, Cham
  (2019)

\end{thebibliography}
\end{document}